\documentstyle[aps,twocolumn,psfig,epsf]{revtex}
\input epsf

\tightenlines

\begin{document}

\title{\Large{Screening of a Moving Parton in the Quark-Gluon Plasma}}

\author{Munshi G. Mustafa$^1$,  Markus H. Thoma$^2$, and Purnendu 
Chakraborty$^1$}

\address{$^1$Theory Group, Saha Institute of Nuclear Physics, 1/AF Bidhan 
Nagar, Kolkata 700 064, India}

\address{$^2$ Centre for Interdisciplinary Plasma Science, 
Max-Planck-Institut f\"ur extraterrestrische Physik,
P.O. Box 1312, 85741 Garching, Germany}

\maketitle

\vspace{0.4in}

\begin{abstract}
The screening potential of a parton moving through a quark-gluon plasma is 
calculated using the semi-classical transport theory. An anisotropic 
potential showing a minimum in the direction of the parton velocity is 
found. As consequences possible new bound states in the quark-gluon plasma
and $J/\psi$ dissociation are discussed.
\end{abstract}

\vspace{0.2in}

Screening of charges in a plasma is one of the most important collective 
effects in plasma physics. In the classical limit in an isotropic 
and homogeneous plasma the screening potential of a point-like
test charge $Q$ at rest can be derived from the 
linearized Poisson equation, resulting in Debye screening. In this way
the Coulomb potential of a charge in the plasma is modified into a
Debye-H\"uckel or Yukawa potential \cite{LL}
\begin{equation}
\phi (r) = \frac{Q}{r} \exp (-m_D r)
\label{e1}
\end{equation}
with the Debye mass (inverse screening length) $m_D$ ($\hbar = c = k_B =1$). 

A special kind of plasma is the so-called quark-gluon plasma (QGP), where the 
electric charges in a plasma are replaced by the color charges of quarks 
and gluons, mediating the strong interactions between them. Such a state of 
matter is expected to exist at extreme temperatures, above 150 MeV, or 
densities, above about 10 times nuclear density. These conditions could be
fulfilled in the early Universe for the first few microseconds or in the 
interior of neutron stars. In accelerator experiments high-energy 
nucleus-nucleus collisions are used to search for the QGP. In these
collisions a hot and dense fireball is created which might consist of a QGP 
in an early stage (less than about 10 fm/c) \cite{Mueller2}. Since the masses
of the lightest quarks and of the actually massless gluons are much less 
than the temperature 
of the system, the QGP is an ultrarelativistic plasma. To achieve a 
theoretical understanding of the QGP, methods from quantum field theory (QCD)
at finite temperature are adopted \cite{Kapusta}. 
Perturbative QCD should work at 
high temperatures far above the phase transition where the interaction between
the quarks and gluons becomes weak due to a specific property of QCD called
asymptotic freedom. An important quantity which can be derived in this way 
is the polarization tensor describing the behavior of interacting
gluons in the QGP. From the polarization tensor important properties of the 
QGP, such as the dispersion relation and damping of the plasma modes
or the Debye screening of color charges in the QGP can be derived 
\cite{Kajantie}.  

In the QGP the Debye mass of a chromoelectric charge follows
from the static limit of the longitudinal polarization tensor in the 
high-temperature limit \cite{Kajantie},
\begin{equation}
\Pi_{00} (\omega =0, k) = -m_D^2 = -g^2T^2 \left (1+\frac{n_f}{6}\right ),
\label{e2} 
\end{equation}
where $g$ is the strong coupling constant and  $n_f$ the number of light 
quark flavors in the QGP with $m_q \ll T$.

The high-temperature limit of the polarization tensor corresponds 
to the classical approximation. For instance, it is closely related
to the dielectric function following from the semi-classical
Vlasov equation describing a collisionless plasma.
E.g., the longitudinal dielectric function following from 
the Vlasov equation is given by \cite{Silin,Klimov,Weldon}
\begin{eqnarray}
\epsilon_l (\omega ,k) &=& 1-\frac{\Pi_{00}(\omega ,k)}{k^2}\nonumber \\
&=& 1+\frac{m_D^2}{k^2} \left (1-\frac{\omega}{2k} \ln \frac {\omega +k}
{\omega -k}\right ),
\label{e3}
\end{eqnarray}
(The only non-classical inputs here are 
Fermi and Bose distributions instead of the Boltzmann distribution.) 
Quantum effects in the Debye mass have been considered using the 
hard-thermal-loop resummation scheme \cite{Rebhan}, dimensional reduction 
\cite{Kajantie2}, and QCD lattice simulations \cite{Karsch}.
The Debye mass and polarization tensor have also been computed in the case 
of an anisotropic QGP \cite{Biro,Mrow,Romatschke}.

The modification of the confinement potential below the critical temperature
into a Yukawa potential above the critical temperature
might have important consequences for the discovery 
of the QGP in relativistic heavy-ion collisions. Bound states of
heavy quarks, in particular the $J/\psi$ meson, which are produced in the
initial hard scattering processes of the collision, will be dissociated in 
the QGP due to screening of the quark potential and break-up by energetic
gluons \cite{Patra}. Hence the suppression
of $J/\psi$ mesons have been proposed as one of the most promising
signatures for the QGP formation \cite{Matsui}. Indeed, a suppression
of $J/\psi$ mesons has been observed experimentally \cite{exper} and
interpreted as a strong indication for the QGP formation in relativistic
heavy-ion collisions \cite{Blaizot}.

In most calculations of the screening potential in the QGP so far, it was 
assumed that the test charge is at rest. However, quarks and gluons
coming from initial hard processes receive a transverse momentum which causes
them to propagate through the QGP \cite{jet}. In addition, hydrodynamical
models predict a radial outward flow in the fireball \cite{flow}. Hence,
it is of great interest to estimate the screening potential of a parton
moving relatively to the QGP. Chu and Matsui \cite{Chu} have used
the Vlasov equation to investigate dynamic Debye screening for a heavy
quark-antiquark pair traversing a quark-gluon plasma. They found that the 
screening potential becomes strongly anisotropic.

In the case of a non-relativistic plasma the 
screening potential of a moving charge $Q$ with velocity $v$ follows from
the linearized Vlasov and Poisson equations as \cite{Ichimaru,Spatschek}
\begin{equation}
\phi({\vec r}, t; {\vec v}) = \frac{Q}{2\pi^2} \int d^3k
\frac{\exp{[-i{\vec k}\cdot ({\vec r}-{\vec v}t)]}}
{k^2 {\rm Re}[\epsilon_l (\omega={\vec k}\cdot {\vec v}, k)]}.
\label{e4}
\end{equation}
It is easy to show that this expression reduces to the Yukawa potential in 
the case of small velocities, $v=|{\vec v}\, |\ll v_{th}$, 
where $v_{th}$ is the thermal 
velocity of the plasma particles. In the opposite case, $v\gg v_{th}$,
the Coulomb potential is recovered since a screening charge cloud cannot be
formed for fast particles.

The above equation (\ref{e4}) also holds in the case of a relativistic plasma.
We only have to use the relativistic expression (\ref{e3}) for the 
longitudinal dielectric function. For small velocities, $v \rightarrow 0$, 
i.e. $\omega \ll k$, we obtain
\begin{equation}
\epsilon_l (\omega \ll k) = 1+\frac{m_D^2}{k^2},
\label{e5}
\end{equation}
from which again the (shifted) Yukawa potential results \cite{Spatschek}
\begin{equation}
\phi({\vec r}, t; {\vec v}) = \frac{Q}{|{\vec r}-{\vec v}t|} \,
\exp (-m_D |{\vec r}-{\vec v}t|).
\label{e6}
\end{equation}
It should be noted that the opposite limit $v\gg v_{th}$, 
leading to a Coulomb potential in the non-relativistic case,
cannot be realized in an ultrarelativistic plasma because the thermal 
velocity of the plasma particles is given by the speed of light, $v_{th}=c=1$.

In the general case, for parton velocities $v$ 
between 0 and 1, we have to 
solve (\ref{e4}) together with (\ref{e3}) numerically. Since the potential
is not isotropic anymore due to the velocity vector ${\vec v}$,
we will restrict ourselves only to two cases, ${\vec r}$ parallel to ${\vec v}$
and ${\vec r}$ perpendicular to ${\vec v}$, i.e., for illustration
we consider the screening potential only in the direction of the moving 
parton or perpendicular to it. 

In Fig.1 the screening potential
$\phi /Q$ in ${\vec v}$-direction is shown as a function of $r'=r-vt$, 
where $r=|{\vec r}|$, between 0 and 6 fm for various velocities. 
For illustration we have chosen a strong
fine structure constant $\alpha_s =g^2/(4\pi)=0.3$, 
a temperature $T=0.25$ GeV, and the number of quark flavors $n_f=2$. 
The shifted potentials depend only on $v$ and
not on $t$ as it should be the case in a homogeneous and isotropic plasma. 
For $r'<1$ fm one observes that the fall-off of the potential
is stronger than for a parton at rest.
The reason for this behavior is the fact that there is a stronger 
screening in the direction 
of the moving parton due to an enhancement of the particle density 
in the rest frame of the moving parton.

In addition, a minimum in the screening potential at $r'>1$ fm shows up. For 
example, for $v=0.8$ this minimum is at about 1.5 fm with a depth 
of about 8 MeV. The occurrence of a minimum in the potential in the 
direction of the velocity is also observed in so-called complex plasmas. 
Complex plasmas are classical, low-temperature plasmas 
containing particles with a diameter of a few microns \cite{Merlino}. These 
particles are charged in the plasma by collecting electrons on their surface.
In the presence of an ion flow the positively charged ions are deflected by the
microparticles, leading to an anisotropic distortion of the Debye sphere by
an enhancement of the ion density in front of the microparticle. This 
positive charge cloud leads to an attraction between microparticles
in the direction of the ion flow and the formation of string like 
structures \cite{Shukla}, observed in experiments.
 
A minimum in the screening potential is also known from non-relativistic, complex
plasmas, where an attractive potential even between equal charges
can be found if the finite extension of the charges is considered \cite{Tsytovich}. 
A similar screening potential was found for a color charge at rest
in Ref.\cite{Gale}, where a polarization tensor beyond the high-temperature limit was used. 
However, this approach has its limitation as a gauge dependent and incomplete (within the order
of the coupling constant) approximation for the polarization tensor was used \cite{Braaten}. 
Obviously, a minimum in the interparticle potential in a relativistic
or non-relativistic plasma is a general feature if one goes beyond the Debye-H\"uckel
approximation by either taking quantum effects, finite velocities, or finite sizes of the particles
into account.
 
Note that Chu and Matsui \cite{Chu} did not report the existence of a minimum in the potential
of a quark traversing the QGP. However, in their Fig.1(d) a negative value of the potential of
a fast quark ($v=0.9$) in the direction of the quark velocity is shown. Since the potential has 
to tend to zero for large distances, this implies a minimum in the screening potential.
This minimum was not found because the potential was plotted only for the limited range
$0<r<1/m_d\simeq 0.35$ fm for our choice of the parameters.

The minimum could give rise to bound states, e.g., of diquarks, if thermal fluctuations do not 
destroy them. The two-body potential, associated with the dipole fields 
created by two test charges $Q_1$ and $Q_2$ at ${\vec r_1}$ and ${\vec r_2}$
with velocities ${\vec v_1}$ and ${\vec v_2}$, can be written as 
\begin{eqnarray}
&&\Phi ({\vec r_1}-{\vec r_2}, {\vec v_1-\vec v_2}, t) = \nonumber \\
&&\frac{Q_1Q_2}{4\pi^2}\> \int d^3k \>
\Biggl \{ \frac{\exp{[i{\vec k}\cdot (-({\vec r_1}-{\vec r_2})-({\vec v_1}-{\vec v_2})t)]}}
{k^2 {\rm Re}[\epsilon_l (\omega={\vec k}\cdot {\vec v_1}, k)]}\nonumber \\
&&+\frac{\exp{[i{\vec k}\cdot (({\vec r_1}-{\vec r_2})-({\vec v_1}-{\vec v_2})t)]}}
{k^2 {\rm Re}[\epsilon_l (\omega={\vec k}\cdot {\vec v_2}, k)]}\Biggr \}.
\label{e7}
\end{eqnarray}
For comoving quarks, ${\vec v_1} = {\vec v_2}$, this two-body potential reduces
to the one-body potential (\ref{e4}), showing the attraction between the quarks
which could give rise to a bound state.
Colored bound states, e.g., diquarks, of partons at rest have also been 
claimed by analyzing lattice
data \cite{Shuryak}. For a quark-antiquark system, where $Q_1Q_2<0$, the two-body potential 
is inverted, showing a maximum. This may lead to short living mesonic resonances 
and an enhancement of the attraction between quarks and antiquarks of mesonic states moving 
through the QGP.   

%%%%%%%%%%%%%%%%%%% figure 1  here %%%%%%%%%%%%%%%%%%%%%%%%%
\vspace{-0.35in}
\begin{figure}
\epsfxsize=3.8in
\hspace{-0.2in}\epsfbox{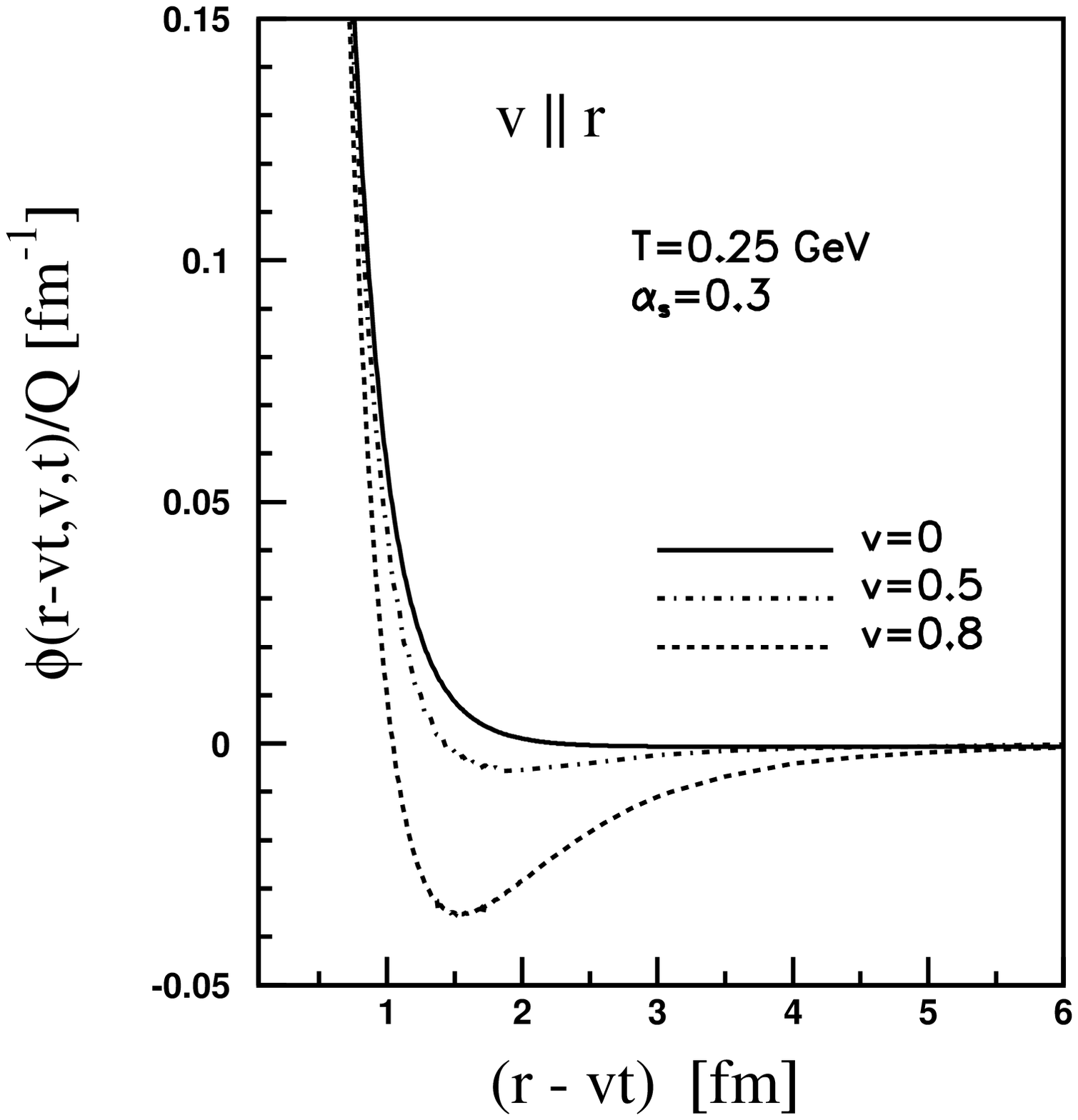}
\vspace{-0.8in}

\noindent{Fig.1: Screening potential parallel to the velocity of the moving
parton in a QGP as a function of $r-vt$ (0 to 6 fm) for $v=0$, 0.5, and 0.8.}
\end{figure}
%%%%%%%%%%%%%%%%%%% end of figure 1 %%%%%%%%%%%%%%%%%%%%%%%%%

The details of the potential, e.g., the depth of the minimum,
depend on the choice of the parameters, such as the coupling constant. For a 
value $\alpha_s =0.3$, as it is typical for the temperature reachable in
heavy-ion collisions, the semi-classical approach corresponding to the
weak coupling limit might not be reliable. Quantum effects and collisions
between plasma particles
are important at those temperatures and will change the dielectric functions and dispersion
relations. Within the transport theoretical approach collisions can be considered,
for example by using the relaxation time approximation \cite{Carrington}. 

Non-abelian effects (beyond color factors, e.g., in the Debye mass)
will be important at realistic temperatures. Unfortunately they cannot be 
treated by the methods used here and are therefore beyond the scope of this work. However, as we 
discussed above, also in a complex plasma which is a strongly coupled plasma as it is also the case for 
the QGP close to
the critical temperature, there is an attraction between the microparticles in the presence
of an ion flow. This appears to be a general feature of weakly as well as strongly coupled plasmas.
Therefore we do not expect a qualitative change of the screening potential due to 
non-perturbative and non-abelian effects. 

The results for ${\vec r}$ perpendicular to ${\vec v}$ are shown in Fig.2,
where the potential is shown as a function of $|{\vec r} - {\vec v}t|=\sqrt{r^2+v^2t^2}$
between 0.1 and 1 fm.
Here we consider only the case $t=0$ since at $t>0$ and $v>0$ there is no singularity
in the potential due to $\sqrt{r^2+v^2t^2}>0$ for all $r$. Hence the potential
is cut-off artificially at small distances if plotted as a function of $\sqrt{r^2+v^2t^2}$.
In contrast to the parallel case (Fig.1)
the fall-off of the potential at larger values of $v$ is less steep, i.e. the screening is
reduced as it is expected since the formation of the screening cloud is suppressed at large 
velocities. Also no minimum in the potential is found.

%%%%%%%%%%%%%%%%%%% figure 2  here %%%%%%%%%%%%%%%%%%%%%%%%%
\vspace{-0.35in}
\begin{figure}
\epsfxsize=3.8in
\hspace{-0.2in}\epsfbox{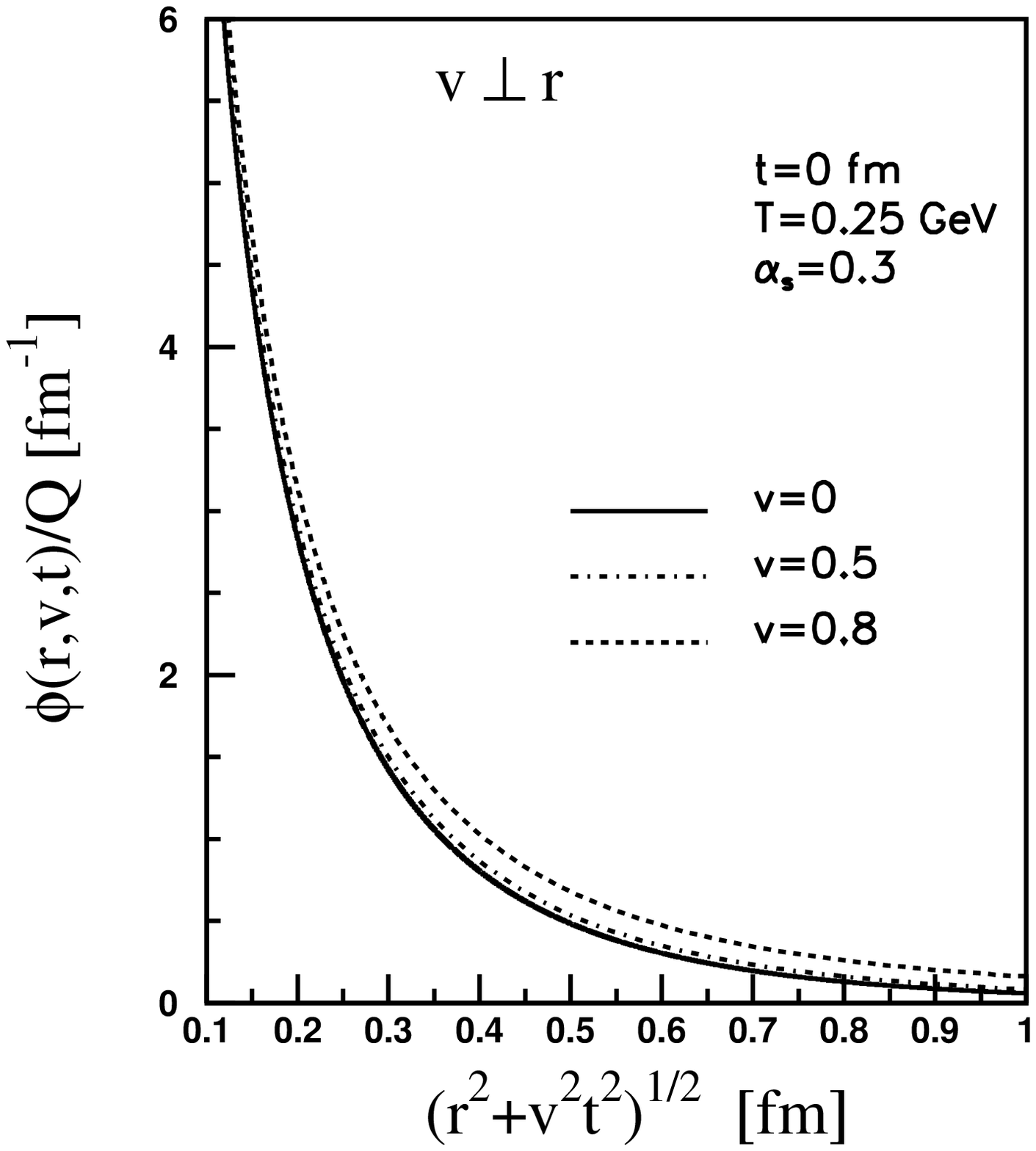}
\vspace{-0.8in}

\noindent{Fig.2: Screening potential perpendicular to the velocity of the moving
parton in a QGP as a function of $| {\vec r} -{\vec v}t|$ (0.1 to 1 fm) for 
$v=0$, 0.5, and 0.8.}
\end{figure}
%%%%%%%%%%%%%%%%%%% end of figure 2 %%%%%%%%%%%%%%%%%%%%%%%%%

Summarizing, we have calculated the screening potential of a color charge
moving through the QGP from semi-classical transport theory corresponding to the high-temperature 
limit. As in Ref.\cite{Chu} we obtained a strongly anisotropic screening potential. 
The screening is reduced in the direction perpendicular of the moving
parton but increased in the direction of the moving parton, which may lead to a modification 
of the $J/\psi$ suppression. In addition,
we found a new feature of the screening potential of a fast parton in a QGP:
a minimum in the potential shows up which could give rise to bound states of, for example,
diquarks if not destroyed by thermal fluctuations. 
For a quark-antiquark pair this minimum turns into a maximum which could cause short living 
mesonic resonances. Combining the effect of reduced screening in perpendicular direction and
the presence of a maximum in parallel direction we expect a stronger binding of 
moving $J/\psi$ mesons than of $J/\psi$ mesons at rest with respect to the QGP.
The consequences, e.g, 
for the $J/\psi$ yield should be investigated in more detail using hydrodynamical models or
event generators for the space-time evolution of the fireball. Finally, let us note that our results also 
applies to other ultrarelativistic plasmas such as an electron-positron plasma in Supernova explosions.
In this case one simply has to replace the Debye mass $m_D$ by $eT$, where $e$ is the electron charge.  

\vspace{0.15in}

\noindent{\it Acknowledgment:} M.H.T. would like to thank B. Klumov and 
V.N. Tsytovich for useful 
discussions and M.G.M is thankful to S. Mallik, D. Mukhopadhyay,
and D. K. Srivastava for useful discussions.

\end{document}